\documentclass{aastex62}

\graphicspath{{./}{figures/}}

\received{...}
\revised{...}
\accepted{...}

\submitjournal{ApJ}

\shorttitle{ }
\shortauthors{Yang et al.}

\begin{document}
\title{How does the Earth's rotation affect predictions of gravitational wave strong lensing rates?}

%\correspondingauthor{Lilan Yang}
%\email{yang_lilan@mail.bnu.edu.cn}

\author{Lilan Yang}
\affil{School of Physics and Technology, Wuhan University, Wuhan 430072, China; $yang\_lilan@whu.edu.cn$; $zhuzh@whu.edu.cn$}
\author{Xuheng Ding}
\affil{School of Physics and Technology, Wuhan University, Wuhan 430072, China; $yang\_lilan@whu.edu.cn$; $zhuzh@whu.edu.cn$}

\author{Marek Biesiada}
\affiliation{Department of Astronomy, Beijing Normal University, Beijing, 100875, China; $zhuzh@bnu.edu.cn$}
\affiliation{Department of Astrophysics and Cosmology, Institute of Physics, University of Silesia, 75 Pu{\l}ku Piechoty 1, 41-500, Chorz{\'o}w , Poland}

\author{Kai Liao }
%\altaffiliation{Creator of AASTeX v6.2}
\affiliation{School of Science, Wuhan University of Technology, Wuhan 430070, China;}
%\collaboration{(LaTeX collaboration)}
\author{Zong-Hong Zhu }
%\altaffiliation{Creator of AASTeX v6.2}
\affil{School of Physics and Technology, Wuhan University, Wuhan 430072, China; $yang\_lilan@whu.edu.cn$; $zhuzh@whu.edu.cn$}
\affiliation{Department of Astronomy, Beijing Normal University, Beijing, 100875, China; $zhuzh@bnu.edu.cn$}

\begin{abstract}
The next generation of ground-based gravitational wave (GW) detectors, e.g. the Einstein Telescope, is expected to observe a significant number of  strongly lensed GW events as predicted in many previous papers.
However, all these works ignored the impact of the Earth's rotation on this prediction.
Multiple lensed images arrive at the Earth at different time, thus the ground-based detector has different responses to the lensed images due to different orientations of the detector relative to the GW source direction. Therefore the amplitudes of the GW signal from different images are modulated appropriately, in addition to the lensing magnification.
In order to assess this effect, we performed Monte Carlo simulations to calculate the event rate of lensed GW signals.
%To compare with previous work, we follow \cite{ET2, ET3} and adopt the intrinsic merger rate of double compact object (NS-NS, BH-NS, BH-BH systems) by \cite{Dominik13}.
Our conclusion is that the Earth's rotation has a non-negligible impact on the event rate of lensed GW image. The updated event rates decrease by factors of $\sim40\%, \sim20\%, \sim10\%$, for NS-NS, BH-NS, BH-BH systems respectively.
%However, since the event rate is dominated by BH-BH, the total rate decrease by a factor of $\sim10\%$ which indicate the rotation effect do not have significant impact on the cosmological inference.

\end{abstract}

\keywords{gravitational lensing: strong, gravitational waves }

\section{Introduction}

After the first detection of gravitational wave (GW) event GW150914, which was produced by merging binary black holes (BBH)\citep{Abbott1} and other subsequent detections \cite{Abbott2, Abbott3, GW170814, GW170817,GW170608}, GW astronomy came into being and a brand new window on the Universe was opened.
Especially the event GW170817 registered by the LIGO-Virgo gravitational wave detector network \citep{Abbott3} accompanied by the detection of electromagnetic counterpart was a breakthrough that commenced a new era in the multi-messenger astronomy creating unique opportunities to deepen our understanding of the Universe. One may expect that LIGO-Virgo network will keep providing new detections and the next generation of ground-based detectors such as the Einstein Telescope (ET) operating with increased sensitivities will eventually yield $10^3-10^7$ inspiral events per year reaching the redshift $z=17$ \citep{Abernathy2011}.
Reaching so deep, one may expect that a significant number of such signals could be lensed by intervening galaxies.

Gravitational lensing of GWs has been extensively studied in many works since the pioneering paper \citep{Wang96}.
In particular, the effect of lensing on the parameter extraction of gravitational wave signals was discussed by \citep{Cao2014}.
It was also suggested that cosmological parameters can be significantly constrained
using time delays measurements of strongly lensed GW events  \citep{Sereno2010, Liaokai2017}.
Moreover, strongly lensed GW signals can be used to test fundamental physics. For example, the speed of gravity can be tested with strongly lensed GW events accompanied by electromagnetic counterparts \citep{Fan2017, Collett2017}.
Admittedly, constraints on the speed of GWs (expressed as bounds on the graviton mass or equivalently on its Compton wavelength) obtained by the LIGO Collaboration \cite{LIGOtest}  with un-lensed events are already very strigent. However, they strongly rely on the PPN waveform templates fitted to the data. The idea here is that if the Compton wavelength of the graviton was finite,
lower frequencies would propagate slower than higher frequencies, leading to the dispersion which would modify phasing of the coalescing signal. Similarly, the bound obatined from the coalescing NS-NS system  \cite{2017Abbott} was a conservative one assuming that GW and EM signals were emitted simultaneously and the observed delay was attributed solely to the difference of propagation speeds. Lensed GW-EM signals are free from such pre-assumptions. 
Therefore, the accurate prediction of lensed GW event rate becomes an important issue.

Lensed event rates for the ET detector were studied in \citep{ET1,ET2,ET3}. The intrinsic merger rates of the whole class of double compact objects -- DCOs thereafer -- i.e. (NS-NS,BH-NS,BH-BH systems) located at different redshifts were taken from the {\tt StarTrack} population synthesis evolutionary code \citep{Dominik13}. Optical depth for lensing
was calculated using the singular isothermal sphere (SIS) lens model and Schechter-like velocity dispersion distribution in the population of lenses according to \cite{Choi07}.
The general conclusion was that the ET would register about $50$ -- $100$ strongly lensed inspiral events per year. These would be dominated by BH-BH events contributing $91\%$ -- $95\%$  to the total rate, depending on details of evolutionary scenarios considered. Recently, \citet{Li2018} extended these predictions to more realistic lens properties allowing for the ellipticity of the lens (and thus the quadruple lensed images), the lens  environment (modeled as an external shear), and magnification bias.
Meanwhile, the prediction of the lensing rate for Advanced LIGO was revisited by \citep{Ng2018}, where it had been noticed that the detection is correlated with source position and detector's orientation.
Comparing to ground based detectors,  lensing rate for the space-borne detector LISA were also discussed in  \citep{Sereno2010, Sereno2011}.

All these works done so far, ignored the impact of the Earth's rotation. The orientation between the source and the detector was considered as  fixed. However, since different lensed signals travel along different paths and probe different depths of the potential of the lens (Shapiro effect), they actually arrive at the Earth at different times. The typical time delay of multiple signals varies from days to hundred of days. After a time delay, the rotation of the Earth changes the orientation of the ground-based detector with respect to the direction of the GW source.  As will see, this will affect the strength of the observed signal, in addition to the relative magnification due to lensing. If not taken into account, it would bias the predictions.

In this paper, we are filling this gap by accounting for the Earth rotation in estimating the rates of lensed GW signals. We base our forecasts on the Monte Carlo simulation. In section \ref{method}, we describe our methodology. In section  \ref{results&dis}, we present and discuss the results. Finally, the conclusions are summarized in section \ref{conclusion}.

\section{Methodology}\label{method}

Our predictions are focused on the ET, which will consist of three nested detectors placed underground at depth of 100 - 200 m, arranged in a triangular pattern. Initial design assumed that each detector would be built from a single interferometer, where the high power needed to achieve good high-frequency performance compromises the low-frequency performance. The next step in the ET design is the so called ``xylophone configuration'', where each detector is split into two interferometers, one specialised for detecting low-frequency gravitational waves and the other one for the high-frequency part. 

We briefly review the detection rate for unlensed events in section \ref{unlensed_detection} and lensing statics in section  \ref{lens_statistics}. Then, we describe the details of our Monte Carlo simulation and update the event rate of lensed GW sources that could be  detected by the ET in section \ref{MC_approach}.

\subsection{Detection rate of unlensed events}\label{unlensed_detection}

Since both the rudiments and the details of GW detection theory have been introduced many times \citep{Finn93, Taylor12, ET1, ET2}, we only recap the main points for clarity. The matched filtering is a standard technique applied in GW data-analysis to efficiently search for GW signals with known characteristics (templates) hidden in noisy data. Consequently, the strength of the signal is measured by the signal-to-noise ratio (SNR).
The optimal matched-filtering SNR of an inspiraling DCO system at the redshift $z_s$ registered by a single detector is
 \begin{equation} \label{SNR}
 \rho = 8 \Theta \frac{r_0}{d_L(z_s)} \left(\frac{{\cal M}_z}{1.2\;M_{\odot}}\right)^{5/6}
 \sqrt{\zeta(f_{max})},
 \end{equation}
where: $d_L$ is the luminosity distance to the inspiralling DCO, $r_0$ is detector's characteristic distance parameter. For two configurations considered here, i.e. the ET initial design and the  advanced ``xylophone'' design, $r_0 = 1527 Mpc$  and $r_0 = 1918 Mpc$,  respectively. ${\cal M}_z$ is the observed (redshifted)  chirp mass (i.e. ${\cal M}_z={\cal M}_0(1+z)$) . We have assumed that ${\cal M}_0$ are 1.2 $M_{\odot}$ for NS-NS, 3.2 $M_{\odot}$ for BH-NS and 6.7 $M_{\odot}$ for BH-BH systems respectively. According to \cite{Dominik2012}, these values represent average chirp mass for each category of DCO simulated by population synthesis. They were also used in \cite{Dominik13}. $\zeta(f_{max})$ is the dimensionless function reflecting the overlap between the GW signal and the detector's effective bandwidth. 
 $\Theta$ is the orientation factor determined by four angles,
according to:
 \begin{equation} \label{Theta}
 \Theta = 2 [ F_{+}^2(1 + \cos^2{\iota} )^2 + 4 F_{\times}^2 \cos^2{\iota} ]^{1/2}
 \end{equation}
where: $F_{+} = \frac{1}{2} (1 + \cos^2{\theta}) \cos{2\phi} \cos{2 \psi} - \cos{\theta} \sin{2 \phi} \sin{ 2 \psi}$, and
$F_{\times} = \frac{1}{2} (1 + \cos^2{\theta})  \cos{2\phi} \sin{2 \psi} + \cos{\theta} \sin{2 \phi} \cos{ 2 \psi}$ are called antenna patterns.
Two angles $(\theta ,\phi)$ describe the direction to the DCO binary relative to the detector; another two angles $( \psi,\iota)$ describe the binary orientation relative to the line-of-sight between it and the detector.
These four angles $(\theta ,\phi, \psi,\iota)$ are independent and one can assume that $(\cos\theta, \phi/\pi, \psi/\pi, \cos\iota)$ are distributed uniformly over the range $[-1, 1]$.
Due to rotation of the Earth, $(\theta ,\phi)$ of a given source would change appropriately, while $( \psi,\iota)$ would be the same.  Note that different values of $(\theta ,\phi)$ could modify the value of $\Theta$ up to an order of magnitude, thus the effect of rotation of the Earth is non-trivial. For this reason, $\rho$ of lensed images is not constant but is a function of time $\rho(t)$. Single inspiral GW signal crosses the detector at one well defined moment $t_0$. SNR of such event can be considered as a single value $\rho(t_0)$.
GW signal is detectable, if its SNR exceeds the threshold, i.e. $\rho > \rho_0 = 8$.
Then, the yearly detection rate of DCO sources can be expressed as:
 \begin{equation} \label{unlensed_rates}
{\dot N}(>\rho_0|z_s) = \int_0^{z_s} \frac{d {\dot N}(>\rho_0)}{dz} dz
 \end{equation}
where $\frac{d\dot{N}(>\rho_0)}{dz_{s}}$ is the yearly merging rate of DCO sources in the redshift interval $[z_{s}, z_{s}+dz_{s}]$.
In one such redshift interval, the rate of DCO events is
 \begin{equation}
 d\dot{N}=4\pi\left(\frac{c}{H_{0}}\right)^3\frac{\dot{n}_{0}(z_{s})}{1+z_{s}}\frac{\tilde{r}^2(z_{s})}{E(z_{s})}dz_{s}
 \end{equation}
where $\dot{n}_{0}(z_{s})$ denotes intrinsic inspiral rate at redshift $z_s$, $\tilde{r}(z_{s})$ is dimensionless comoving distance to the source and $E(z_s)$ - dimensionless expansion rate of the Universe at redshift $z_s$. Concerning cosmological model, we assume flat $\Lambda$CDM model with $\Omega_m=0.3$ and $H_0 = 70 \;km\;s^{-1}\;Mpc^{-1}$ in order to comply with \citep{Dominik13, ET2, ET3}. 
More importantly, we use the intrinsic merger rates $\dot{n}_{0}(z_{s})$ predicted by the population synthesis model (using {\tt StarTrack} code) in \citet{Dominik13}. We have taken the
data from the website http://www.syntheticuniverse.org, more specifically the so called ``rest frame rates'' in cosmological scenario. In this code  binary systems were evolved from ZAMS until the compact binary formation (after supernova explosions) making a number of assumptions about star formation rate, galaxy mass distribution, stellar populations, their metallicities and galaxy metallicity evolution with redshift . 
In order to investigate the
uncertainties of the chemical evolution of the Universe \citep{Dominik13} employed two distinct scenarios for metallicity evolution with redshift, called ``low-end'' and ``high-end''scenarios.
These assumptions result with different DCO merger rates predictions and consequently to distinct lensed GW event rates. 
Because the compact object formation depends critically on the physics of common envelope (CE) phase of evolution and on SN explosion mechanism and both of them are to some degree uncertain, \citet{Dominik13} considered four scenarios: standard one and three of its modifications --- Optimistic Common Envelope (OCE), delayed SN explosion and high BH kicks scenario. Some of the underlying assumptions could be constrained in light of existing GW detections \citep{Chruslinska18}. However, the aim of this work is to evaluate the effect of the Earth's rotation, hence we focus only on the standard scenario to facilitate the comparison with previous works.

\subsection{Lensing statistics}
\label{lens_statistics}
Similarly as in \citep{ET2,ET3}, we assume the SIS model of the lens. Of course, adopting more sophisticated lens models (such as SIE or power-law profile) could make the prediction more realistic and would enable to study lensing systems with quadruple images. In the recent paper \cite{Li2018}, whose authors used the SIE model, the results turned out to be consistent with the ones by SIS model (\cite{ET2, ET3}), thus we conclude that the SIS model is sufficient for the purpose of our study, which is the prediction of the event rates.
From the physical point of view, mass distribution of lensing galaxies is the most relevant parameter for gravitational lensing.  
In the SIS model assumed here, the same information is conveyed by the  stellar velocity dispersion, which is much easier to assess.

Einstein radius $\theta_{E}$ gives the characteristic angular scale of lensing phenomenon. In the SIS model, it can be expressed  as $\theta_{E}=4\pi(\frac{\sigma}{c})^2\frac{d_{A}(z_{l},z_{s})}{d_{A}(z_{s})}$, where $\sigma$ denotes velocity dispersion of lensing galaxy, $d_{A}(z_{s})$ and  $d_{A}(z_{l},z_{s})$ are angular diameter distances to the source and between the source and the lens, respectively. Introducing the angle $\beta$, which is the angle between directions to the source location and to the center of the lens, one
concludes that if $\beta<\theta_{E}$, two strong lensed images (brighter one $I_{+}$ and fainter one $I_{-}$) will appear at locations $\theta_{\pm}=\theta_{E}\pm\beta$. Introducing dimensionless quantities:  $x=\frac{\theta}{\theta_{E}}$ and $y=\frac{\beta}{\theta_{E}}$, the strong lensing condition becomes $y<1$ and positions of the images are $x_{\pm}=1\pm y$ with magnifications $\mu_{\pm}=\frac{1}{y}\pm1$. Gravitationally lensed GW signals, corresponding to these two images would have SNRs: $\rho_{\pm}=\rho_{intr._\pm}\sqrt{\mu_{\pm}}=\rho_{intr._\pm}\sqrt{\frac{1}{y}\pm1}$. Let us emphasize that the $\rho_{intr._-}$ and $\rho_{intr._+}$ are now different due to the rotation of the Earth.

In previous works, only the $I_{-}$ image was considered necessary for the detection in order to establish the lensed nature of the signal.
It was because its SNR $\rho_- = \rho_{intr._-}\sqrt{\frac{1}{y}-1}$ was mistakenly assumed as being always lower than the SNR of $I_+$ i.e.  $\rho_+ =\rho_{intr._+}\sqrt{\frac{1}{y}+1}$, which lead to a misconception that detecting the $I_-$ ensures the ability to detect $I_+$. However, due to  the rotation of the Earth during the time delay between images it need not be the case. Therefore, we should treat the $I_-$ and $I_+$ images  separately and we propose that the lensed GW event registered by the ET requires both $\rho_-$ and $\rho_+$ exceeding the threshold $\rho_0 = 8$. This means that $\sqrt{\frac{1}{y}\pm1}\rho_{intr._\pm}>\rho_0$ is required, which leads to the condition:
\begin{equation} \label{y_condition}
y_{\pm} \leq y_{\pm,max} = \left[ \left( \frac{8}{\rho_{intr.{\pm} }} \right)^2 \mp 1 \right]^{-1}
\end{equation}
Thus, the elementary cross section for lensing reads:
\begin{equation} \label{cross_section}
S_{cr, \pm}(\sigma, z_l, z_s, \rho_{\pm}) = \pi \theta_E^2 y_{\pm,max}^2  = 16 \pi^3 \left( \frac{\sigma}{c} \right)^4 \left( \frac{{\tilde r}_{ls}} {{\tilde r}_{s}} \right)^2 y_{\pm, max}^2 .
\end{equation}
Finally, total optical depth $\tau$ describes probability that GW source at redshift $z_{s}$ would be lensed  and detected by the ET.
Since the images $I_{+}$ and $I_{-}$ are considered separately, we split $\tau$ into two terms, i.e. $\tau_{\pm}$ corresponding to $I_\pm$, respectively:
\begin{equation} \label{tau}
\tau_{\pm}(z_s, \rho_{\pm}) = \frac{1}{4 \pi} \int_0^{z_s}\; dz_l \; \int^{\infty}_0 \; d \sigma \; %\frac{dn}{d \sigma} S_{cr}(\sigma, z_l, z_s) \frac{d V}{d z_l}
4 \pi \left( \frac{c}{H_0} \right)^3 \frac{ {\tilde r}_l^2}{E(z_l)} S_{cr, \pm}(\sigma, z_l, z_s, \rho_{\pm}) \frac{d n}{d \sigma}
\end{equation}
where $z_{l}$ is the redshift of lens (we assume that lenses are distributed homogeneously in redshift). Moreover, 
we model the velocity dispersion distribution in the population of lensing galaxies as a modified Schechter function $\frac{d n}{d \sigma} = n_{*} \left( \frac{\sigma}{\sigma_{*}} \right)^{\alpha} \exp{\left( - \left( \frac{\sigma}{\sigma_{*}} \right)^{\beta} \right)} \frac{\beta}{\Gamma (\frac{\alpha}{\beta}) } \frac{1}{\sigma}$, with the parameters $n_{*}$,$\sigma_{*}$,${\alpha}$ and $\beta$  taken after \citet{Choi07}. This is the same assumption as made in \cite{ET3} - we reproduced it for comparison of results. 
Admittedly, there exist more recent data for the velocity distributions of galaxies like e.g. \citet{Bernardi2010}. However,  these fits were made for galaxies of all types, unlike  \citep{Choi07} who used early type galaxies only.

\subsection{Monte Carlo method}
\label{MC_approach}

Our goal is to update the event rates of lensed GW signals, explicitly considering variation of the
detector's orientation due to Earth's rotation, which will modulate the SNRs of $I_-$and $I_+$. This can hardly be done analytically, hence we need to perform the appropriate Monte Carlo simulation.
We describe the details of the simulation in this section.

\begin{figure}
\begin{center}
\includegraphics[width=80mm]{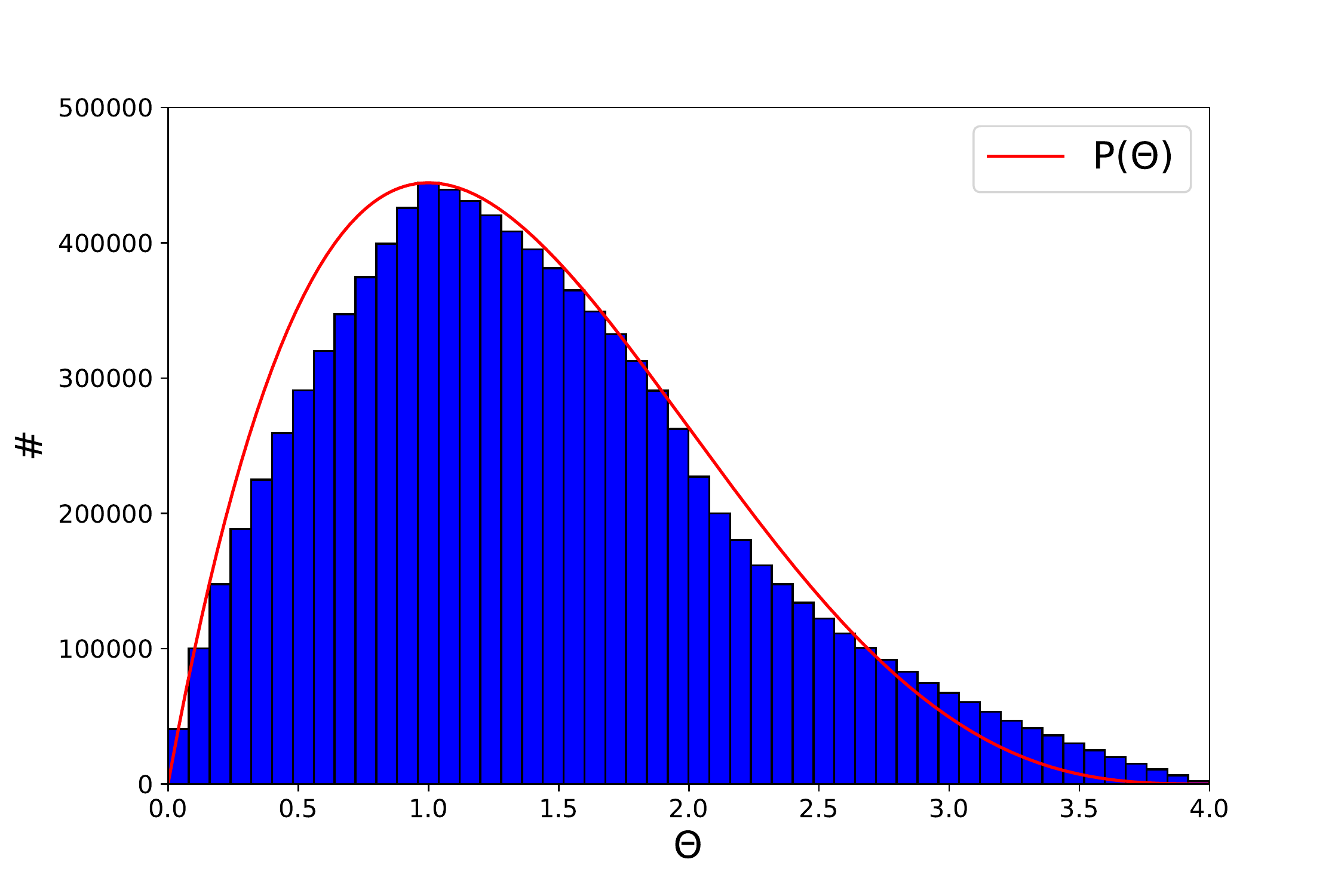}
\end{center}
\caption{Histogram of $\Theta$. The red line shows numerical probability distribution of  $\Theta$ Eq.(\ref{P_theta})}
\label{Thetadis}
\end{figure}

We build up a mock universe by creating a large sample of DCO events to represent the overall DCO events, and randomly generate the values of their key parameters. For a DCO event at redshift $z_s$, its $\rho$ at a specific moment can be randomly assigned with Eq.(\ref{SNR}). This is done by randomly assigning the orientation factor $\Theta$ based on sampling the four angles $(\theta ,\phi, \psi,\iota)$\footnote{$(\cos\theta, \phi/\pi, \psi/\pi, \cos\iota)$ are distributed uniformly in the  range [-1, 1].}. Let us note, that alternatively to the Monte Carlo method, the common practice is to define the $\Theta$ by taking its averaged numerical probability density as (see formula 3.11 in \cite{Finn1996}):
\begin{eqnarray} \label{P_theta}
P_{\Theta}(\Theta) &=& 5 \Theta (4 - \Theta)^3 /256, \qquad {\rm if}\;\;\;
0< \Theta < 4  \\
P_{\Theta}(\Theta) &=& 0, \qquad {\rm otherwise.} \nonumber
\end{eqnarray}\\
It is instructive to recall the origin of the Eq(\ref{P_theta}). First,  \cite{Finn93} numerically estimated $P(\Theta^2)$ using Monte Carlo simulation -- see Table 1
in \cite{Finn93}, where cumulative probability distribution of $\Theta^2$ is reported. Later, 
\cite{Finn1996} noticed that ``to an excelent approximation'' $P_{\Theta}(\Theta)$ can be described by the Eq(\ref{P_theta}). Since then this equation has been used abundantly due to its simple analytical form. 
In Figure \ref{Thetadis} we compare our random sampling of $\Theta$ based on $10^7$ sample points (the same number as used by \cite{Finn93}) to this numerical probability density. 
One can see, that the numerical probability distribution Eq(\ref{P_theta}) is not an excellent approximation.  

Identifying lensed DCO system by the ET requires that both lensed images  $I_-$ and $I_+$ are detected.
We break this identification in two steps.
We start with answering how many lensed DCO events have $I_-$ available to the ET and not considering $I_+$ for a moment, and the corresponding optical depth is $\tau_{-}(z_s,\rho_{intr._-})$ (see Eq(\ref{tau})).
Knowing this probability, the total number of $I_-$ detected  can be sampled by Monte Carlo simulation by accumulating the events which meet the requirement though all the space, i.e., $ \int_0^{z_{max}} \int_0^{\infty} d\rho \tau(z_s,\rho) \frac{\partial {\dot N}}{\partial z\partial \rho}  dz$.
Let us remark that quantitatively this number should be equivalent to numerical values reported in previous papers \cite{ET2,ET3}.

\begin{figure}
\begin{center}
\includegraphics[width=80mm]{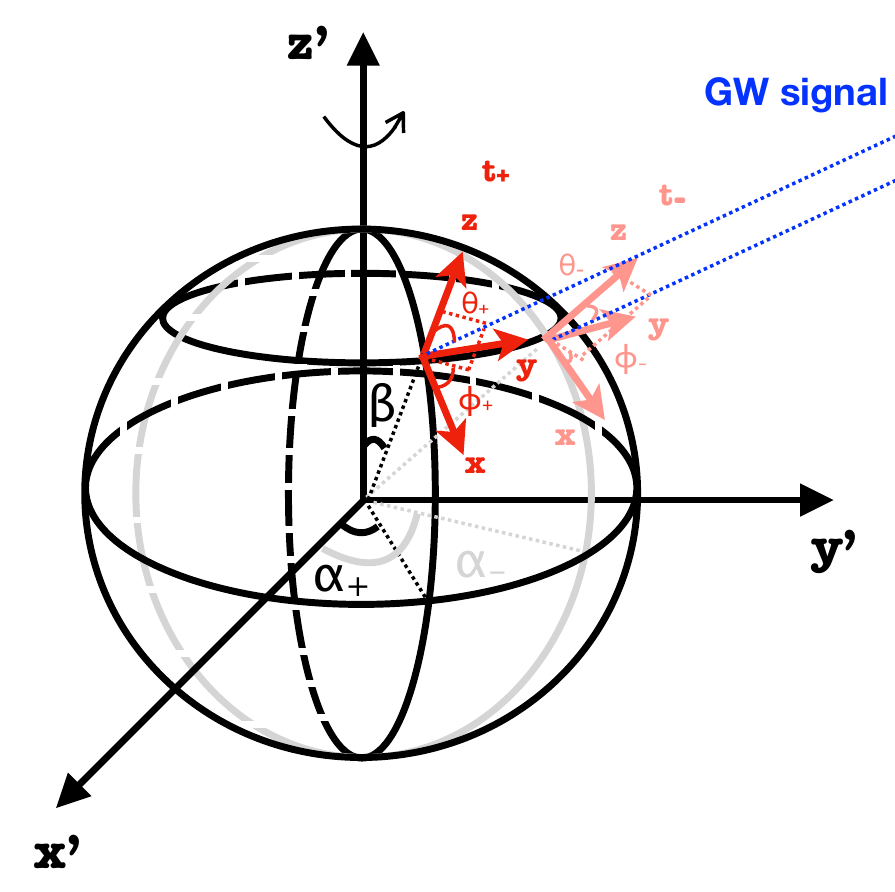}
\end{center}
\caption{Sketched map of regenerating the values of $(\theta_+,\phi_+)$. The red and pink coordinate systems represent the coordinate systems fixed on the ET at $t_+$ and $t_-$, respectively.  The angles $(\beta,\alpha)$ describe the location of the ET on the earth coordinate system. The  $\beta$ is kept constant while the $\alpha$ varies with the earth's rotation. The blue dotted line represents the direction of the GW signal. }
\label{rotation}
\end{figure}

Once the set of ``$I_-$ detected'' events is sampled, we can randomly assign the position $y$ for each of them\footnote{The square of source position, i.e., $y^2$ follows a uniform distribution between $(0, y_{max}^2$).}. The SNR for corresponding $I_+$ images can thus be calculated as $\rho_+ = \rho_{intr._+}\sqrt{\frac{1}{y}+1}$. As mentioned above, $\rho_{intr._+}$ is different from $\rho_{intr._-}$ due to time delay. To take this difference into account, we keep the $(\psi,\iota)$ values unchanged and re-generate the values of $(\theta ,\phi)$ to derive the $\rho_{intr._+}$.  This re-generating depends on the location of the ET on Earth and on the time delay between  ``$I_-$'' and  ``$I_+$''.
%We describe the details of re-generating process in the Appendix \ref{append}. 
In order to re-generate the values of $(\theta_+, \phi_+)$ at $t_+$, knowing their initial values $(\theta_-, \phi_-)$ at $t_-$, where $t_+$ and $t_-$ are the times of arrival for $I_+$ and $I_-$, respectively\footnote{Let us remind that $t_+ = t_- - \Delta t$, where $\Delta t$ is lensing time delay.}, we proceed in the following way. The angles $(\theta, \phi)$ are defined in a local detector's coordinate system, see Figure~\ref{rotation}. For simplicity we assume that it is a cardinal coordinate system of the ET located at latitude $45^\circ N$, longitude $0^\circ$ \citep{Abernathy2011}. Then we transform the coordinate of the GW signal to the center of the Earth with $z$ axis coincident with the Earth's rotation axis. 
In this new non-rotating coordinate system, the direction of the GW
signal is fixed \footnote{We neglect the orbital motion of the Earth
and corresponding change
of direction to the GW source.}.
The angles $(\beta, \alpha)$ are used to describe the location of the ET: 
$\beta=const.=45^\circ$, while $\alpha = \alpha_0 + \Omega_r t$, where $\Omega_r = \frac{2 \pi}{T}$ is the rotational angular velocity of the Earth. 
%β =const: = 45◦, while
%α = α0 + Ωrt, where Ωr = 2 T π is the rotational angular velocity of
%the Earth. 
Consequently, 
$\alpha_+ = \alpha_- - \Omega_r \Delta t$,
and knowing the time delay $\Delta t$, we can transform the coordinate of the
GW signal back to the detector's frame. Furthermore, we assume the time delay $\Delta t$ follows a uniform distribution as [0, 24hr]. Of course what matters here is $\Delta t \bmod 24hr$.

Knowing $\rho_{intr._+}$, we obtain the optical depth $\tau_{+}(z_s,\rho_{intr._+})$ for ``$I_+$" which enables one to count the number of events having $\rho_+>\rho_0 = 8$ across all the redshift bins. Finally, we count the events for which both the $I_-$ and $I_+$ could be detected by the ET.
Note that our procedure of considering the lensed image of $I_-$ first  and then $I_+$ does not mean that the $I_-$ arrives earlier than $I_+$. Actually, the $I_+$ arrives first. However, the identification of lensed GW requires that both $I_-$ and $I_+$ are detected. In this work, we consider detection of $I_-$ as the prior condition.
%This strategy facilitates the comparison with the previous work which only considered the $I_-$ image. Furthermore, since the numbers of the $I_- $ images are much less than $I_+$ image, by first considering the $I_-$ would reduce the computing time by great amount.

We performed $10^5$ realizations of Monte Carlo simulation for all DCO systems and averaged the results.

\begin{table*}[ht]
\footnotesize %\tiny \scriptsize \footnotesize \small
\caption{Predictions of yearly lensed GW event rates for which only $I_-$ image or both $I_-$ and $I_{+}$ images are magnified above the threshold $\rho_0 = 8$. 	 Results are shown for the standard model of DCO formation and two configurations of the ET. The ``high" and ``low" represent the ``high-end" and ``low-end" galaxy metallicity evolution.  }
\label{lensing-results}
\begin{center}  %\scriptsize%\footnotesize
\begin{tabular}{cccccc}
\hline
%standard DCO scenario  && & & \\
Metallicity Evolution& High&High& Low&Low\\
Hich Event Rate& only $I_-$ &  $I_-$  and  $I_+$ &only $I_-$ & $I_-$ and $I_+$ \\
\hline\\
NS-NS & & & & \\
Initial Design &0.7&0.4& 0.6&0.4\\
Xylophone &1.4&1.1& 1.2&0.7\\
\\
\hline\\
BH-NS & & & & \\
Initial Design &2.2&1.8&2.9&2.3\\
Xylophone &3.5&2.9& 4.3&3.6 \\
\\
\hline\\
BH-BH & & & & \\
Initial Design &106.6&94.3& 130.3&115.4 \\
Xylophone &143.5&128.0& 177.6&159.2\\
\\
\hline\\
TOTAL && & & \\
Initial Design &109.5&96.5&133.8&118.1\\
Xylophone &148.4&132&183.1&163.5\\
\\
\hline\\
%\hline
\end{tabular}\\
\end{center}
\end{table*}

\section{Results and discussion}\label{results&dis}
\begin{figure}
\begin{center}
\includegraphics[width=90mm]{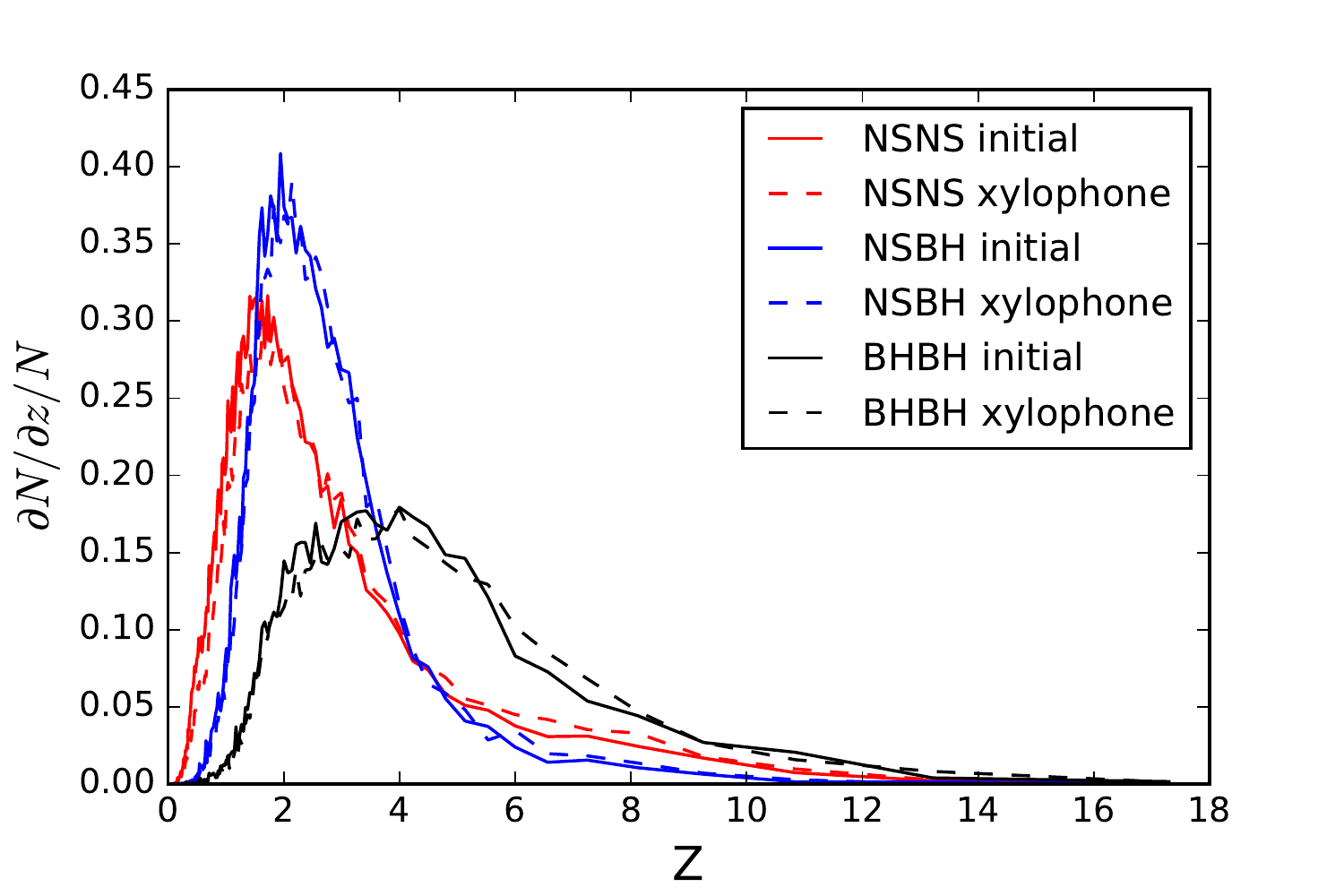}
\end{center}
\caption{Observed lensed GW event number distribution as a function of z.
``Low-end'' metallicity galaxy evolution and standard model of DCO formation are assumed.}
\label{figdis}
\end{figure}

Table~\ref{lensing-results}
shows the  expected yearly rates of lensed GW, based on the standard DCO scenario. Cases when only $I_-$ image is detected and when both $I_-$ and $I_+$ are detected by the ET, are shown.
One can see that the event rate of ``$I_{-}\& I_{+}$" is smaller than the event rate of ``only $I_{-}$". This is the result of rotation of the Earth affecting the prediction of lensed events rates.
Table~\ref{lensing-results} demonstrates that, compared to previous works, our updated event rates decrease by factors of $\sim40\%, \sim20\%, \sim10\%$, for NS-NS, BH-NS and BH-BH systems, respectively. 
The reason why different DCO systems are affected in a distinct way can be understood from the distribution of the intrinsic SNR for different types of binaries (see Fig.~1. of \cite{ET3} ). In the case of NS-NS systems, detectable lensed events (i.e. with $\rho >8$) correspond to the upper tail of the distribution, while the events with $\rho < 8$ dominate in NS-NS systems. Therefore, these systems are more affected by the Earth's rotation than BH-BH systems which are dominated by $\rho > 8$ cases. Threshold value $\rho=8$ splits the distribution of BH-NS systems into two approximately equal parts, which explains the intermediate value of their lensing rate reduction due to rotation of the Earth.
However, since the event rate is dominated by BH-BH systems, total GW lensing rate decreases by a factor of $\sim10\%$. Despite the above described reduction of yearly rates, this means that if the $I_{-}$ image of lensed GW event is detected, then the $I_{+}$ image is very likely to have been detected as well.
One can expect that since the effect of the Earth's rotation has also been neglected in \cite{Li2018, Ng2018}, their prediction of the total lensed event rate are overestimated by $\sim10\%$.
Figure~\ref{figdis} shows the relationship between yearly lensed event rate of ``$I_{-}\& I_{+}$" and the source redshift.
This relationship looks very similar to previous results \citep[Right panel at Fig.~2]{ET3} which indicates that our Monte Carlo simulation achieved  results consistent with numerical calculations.

In this work, we conservatively considered the SIS model, although considering the SIE model (or the power-law profile) would be more realistic and would enable one to study the quadruple lensing systems. However, for the purpose of predicting the lensed GW event rate, SIS model is sufficient, considering fact that the quad fraction is only about $\sim10\%$ of all strong lensing systems \citep{Li2018, Oguri2010}. Of course, the predictions taking into account the Earth's rotation would be more difficult and challenging for the SIE model since there would be three mutual time delays between images. However, the Monte Carlo simulation similar to used in this paper, would then be the only reasonable approach.

The results obtained in this work are applicable to the ground-based detectors only. They are not valid for the next generation space-based detectors, like LISA which will be placed on an Earth-trailing heliocentric orbit \citep{Lisa2017, eLisa2013} with three satellites forming an equilateral triangle. Since the satellites follow their own orbits the triangle will rotate as well, but the rotation period would be about 1 year. This means that the rotation of the LISA detector would have less effect on the lensed GW detection rate than the rotation of the Earth has in the case of ground-based detectors.

\section{conclusion}\label{conclusion}
In this work, we used the Monte Carlo simulation technique to evaluate the effect of the Earth's rotation on the prediction of GWs strong lensing rates for the third generation ground-based detector -- the ET. All previous works concerning GW lensing  ignored this effect and assumed that detecting the fainter image $I_-$  guarantees that the $I_+$ image could be also be observed. This was a  wrong, unjustified assumption. Our results show that the rate of detecting GW signals from both $I_-$ and $I_+$ images is less than the rate of detecting $I_-$ only.
In particular,  for NS-NS systems, the event rate of both $I_-$ and $I_+$ images is $\sim40\%$  less than the rate of detecting only one of them. For BH-BH systems, the impact of the Earth's rotation is the smallest ($\sim10\%$).
Furthermore, we have shown that the Monte Carlo method provides reliable and more accurate results compared D to previous work \citep{ET1, ET2, ET3} using numerical calculation.

Since the total event rate is dominated by BH-BH systems, the impact of the Earth's rotation on the total rate is at the level of $\sim10\%$. Therefore, one should not worry much about this effect making  cosmological inference, such as determination of the Hubble constant or cosmic equation of state, using catalogs of inspiral events. One should keep in mind, however, that this effect is the highest ($\sim40\%$) for  NS-NS systems, which are accompanied by electromagnetic counterparts.
Therefore, the Earth's rotation should be taken into account in all considerations concerning lensed GW signals from coalescing NS binaries.

%\begin{figure}
%\begin{center}
%\includegraphics[width=90mm]{dis_rho.png}
%\end{center}
%\caption{The distribution of $\rho$ for $\rho_{-}$ and $\rho_{+}$ in NS-NS standard low.}
%\label{figdisrho}
%\end{figure}

\acknowledgments
%We thank .....
This work was supported by the National Natural Science Foundation of China under Grants Nos. 11633001 and 11373014, the Strategic Priority Research Program of the Chinese Academy of Sciences, Grant No. XDB23000000 and the Interdiscipline Research Funds of Beijing Normal University.
X. Ding acknowledges support by China Postdoctoral Science Foundation Funded Project (No. 2017M622501).
M.B. expresses his gratitude for hospitality of the Wuhan University where part of this work was done.
M.B. was supported by Foreign Talent
Introducing Project and Special Fund Support of Foreign Knowledge
Introducing Project in China.
K. Liao was supported by the National Natural Science Foundation of China (NSFC) No. 11603015.

%%%%%%%%%%%%%%%%%%%%%%%%%%%%%%%%%%%%%%%%%%%%%%%%%%%%%%%%%%%%%%%%%%%%%%%%%%%%%%%%%%%%%%%%%%%%%%%%%%%%
\bibliography{reference}
\end{document}